\documentclass[prl,aps,twocolumn,groupedaddress]{revtex4}

\usepackage{bm}
\usepackage{graphicx}
\usepackage{color}
\usepackage{ulem}

\begin{document}

\title{Polarity-dependent charge-density wave in a kagome superconductor CsV$_3$Sb$_5$}

\author{Takemi Kato$^{1}$}\thanks{These authors equally contributed to this work.}
\author{Yongkai Li$^{2,3,4}$}\thanks{These authors equally contributed to this work.}
\author{Kosuke Nakayama$^{1,5}$}\email{k.nakayama@arpes.phys.tohoku.ac.jp}
\author{Zhiwei Wang$^{2,3,4}$}\email{zhiweiwang@bit.edu.cn}
\author{Seigo Souma$^{6,7}$}
\author{Miho Kitamura$^{8}$}
\author{Koji Horiba$^{8,9}$}
\author{Hiroshi Kumigashira$^{10}$}
\author{Takashi Takahashi$^{1,7}$}
\author{Takafumi Sato$^{1,6,7,11,}$}\email{t-sato@arpes.phys.tohoku.ac.jp}
\affiliation{$^1$Department of Physics, Graduate School of Science, Tohoku University, Sendai 980-8578, Japan\\
$^2$Centre for Quantum Physics, Key Laboratory of Advanced Optoelectronic Quantum Architecture and Measurement (MOE), School of Physics, Beijing Institute of Technology, Beijing 100081, P. R. China\\
$^3$Beijing Key Lab of Nanophotonics and Ultrafine Optoelectronic Systems, Beijing Institute of Technology, Beijing 100081, P. R. China\\
$^4$Material Science Center, Yangtze Delta Region Academy of Beijing Institute of Technology, Jiaxing, 314011, P. R. China\\
$^5$Precursory Research for Embryonic Science and Technology (PRESTO), Japan Science and Technology Agency (JST), Tokyo, 102-0076, Japan\\
$^6$Center for Spintronics Research Network, Tohoku University, Sendai 980-8577, Japan\\
$^7$Advanced Institute for Materials Research (WPI-AIMR), Tohoku University, Sendai 980-8577, Japan\\
$^8$Institute of Materials Structure Science, High Energy Accelerator Research Organization (KEK), Tsukuba, Ibaraki 305-0801, Japan\\
$^9$National Institutes for Quantum Science and Technology (QST), Sendai 980-8579, Japan\\
$^{10}$Institute of Multidisciplinary Research for Advanced Materials (IMRAM), Tohoku University, Sendai 980-8577, Japan\\
$^{11}$International Center for Synchrotron Radiation Innovation Smart (SRIS), Tohoku University, Sendai 980-8577, Japan\\
}
\date{\today}

\begin{abstract}
Polar surface and interface play a pivotal role for realizing exotic properties of materials, and a search for such polar states is of crucial importance for expanding materials' functionality.
Here we report micro-focused angle-resolved photoemission spectroscopy of CsV$_3$Sb$_5$, a member of recently discovered kagome superconductors AV$_3$Sb$_5$ (A = K, Rb, Cs), and show evidence for the polar nature of cleaved surface which is characterized by Cs- and Sb-terminated surfaces with markedly different fermiology.
The Cs-terminated surface shows intriguing doubling of V-derived bands at low temperature associated with the band folding due to the three-dimensional charge-density wave (CDW), whereas the Sb-terminated one shows no band doubling or resultant CDW-gap opening indicative of the suppression of bulk-originated CDW due to polar charge.
Such polar-surface-dependent band structure must be incorporated for understanding the origin of unconventional superconducting and charge order at the surface of AV$_3$Sb$_5$.
\end{abstract}

\maketitle

Surface and interface of crystals are a fertile playground where a variety of exotic properties emerge owing to its symmetry, chemical-bonding, and topological characteristics distinct from those of bulk counterpart.
Sudden change in the electrostatic potential at the surface generates conducting two-dimensional (2D) electron gas called Shockley states \cite{ShockleyPR1939} which evolve into spin-split Rashba states associated with the space-inversion symmetry breaking \cite{BychkovJETP1984}, whereas the discontinuity of topological index across the surface of topological insulators leads to spin-helical Dirac-cone states \cite{HasanRMP2010,QiRMP2011,AndoJPSJ2013}.
Another key factor that gives a significant influence on physical properties at the surface/interface is polarity; its impact is best demonstrated at the interface of ionic crystals, where the electronic band structure is markedly reconstructed so as to compensate the discontinuity of ionic polarity (known as polar catastrophe).
For instance, LaAlO$_{3}$/SrTiO$_{3}$ interface exhibits high conductivity and even superconductivity despite the semiconducting nature of host crystals \cite{OhtomoNature2004,CavigliaNature2008,ThielScience2006}.
Polarity also plays an essential role at polar surfaces to bring about various exotic properties, as represented by Rashba-split 2D electron gas at SrTiO$_{3}$ surface \cite{MeevasanaNM2011,SantanderNature2011,NakamuraPRL2012,WalkerPRL2014}, extremely over-doped CuO$_{2}$ planes at high-temperature superconductor YBa$_{2}$Cu$_{3}$O$_{7}$ surface \cite{NakayamaPRB2007,ZabolotnyyPRB2007}, ferromagnetic surface of antiferromagntic PdCoO$_{2}$ \cite{MazzolaPNAS2018}, spin-texture manipulation in giant Rashba compound BiTeI \cite{IshizakaNM2011,MaassNCOM2016}, and surface-termination-dependent Fermi arcs in Weyl semimetal NbP \cite{SoumaPRB2016}.
As highlighted by these examples, understanding the physics associated with the polar surfaces of quantum materials is a key topic of modern condensed-matter physics.

Here we focus on $\rm{CsV}_{3}\rm{Sb}_{5}$, a member of kagome metals $\rm{AV}_{3}\rm{Sb}_{5}$ (A = K, Rb, Cs) which were recently discovered to show superconductivity with $T_{\rm{c}}$ = 0.9-2.5 K coexisting with charge-density wave (CDW) below $T_{\rm{CDW}}$ = 78-103 K \cite{OrtizPRM2021,OrtizPRL2020,YinCPL2021} with the in-plane 2$\times$2 periodicity \cite{ChenNature2021,JiangNM2021,ZhaoNature2021,LiangPRX2021,WangPRB2021,ShumiyaPRB2021,NieNature2022,LiarXiv2021}.
The crystal structure of CsV$_3$Sb$_5$ consists of two building blocks, Cs layer and V$_3$Sb$_5$ layer, and the cleaved surface is expected to be terminated either by Cs or V$_3$Sb$_5$ (Sb2) layer [Fig. 1(a)]. Although the bonding between V and Sb1/Sb2 atoms within the V$_3$Sb$_5$ layer may be intermetallic, the coupling between the Cs and V$_3$Sb$_5$ layers is ionic because Cs is fully ionized (Cs$^{1+}$) to donate one electron to the V$_3$Sb$_5$ layer [(V$_3$Sb$_5$)$^{1-}$]. Such ionicity would lead to polar instability.
In fact, two different surface terminations have been identified by scanning-tunneling-microscopy (STM) measurements, and intriguingly, there is growing experimental evidence that such surface hosts peculiar quantum states distinct from those of the bulk, as exemplified by the formation of unidirectional 4$\times$1 CDW \cite{ChenNature2021,ZhaoNature2021,LiangPRX2021,WangPRB2021,ShumiyaPRB2021,NieNature2022,LiarXiv2021,XuPRL2021}, surface-dependent vortex-core states \cite{LiangPRX2021}, and pair-density wave \cite{ChenNature2021}.
However, the band structure associated with the polar surface remains totally unexplored in $\rm{AV}_{3}\rm{Sb}_{5}$.

In this Letter, we report a spatially-resolved angle-resolved photoemission spectroscopy (ARPES) study of $\rm{CsV}_{3}\rm{Sb}_{5}$.
By utilizing the micro-focused photon beam from synchrotron, we have succeeded in separately observing two kinds of polar surfaces, and uncovered marked differences in the band structures between them.
The Cs-terminated surface shows the band doubling associated with the 3D nature of CDW at low temperature, whereas the Sb-terminated surface is strongly hole-doped, resulting in the suppression of CDW.
We discuss implications of the present results in relation to the mechanism of CDW and unconventional surface properties.

High-quality $\rm{CsV}_{3}\rm{Sb}_{5}$ single crystals were synthesized with the self-flux method \cite{OrtizPRM2019}.
ARPES measurements were performed using Scienta-Omicron DA30 and SES2002 spectrometers at BL-28A and BL-2A in Photon Factory, KEK.
We used energy tunable photons of $h\nu$ = 85-350 eV.
ARPES data were mainly obtained by circularly polarized 106-eV photons (corresponding to the out-of-plane wave vector $k_{z}\sim0$) \cite{NakayamaPRB2021} with the beam spot size of 10$\times$12 $\mu$m$^{2}$ \cite{KitamuraRSI2022}.
The energy resolution was set to be 25 meV.
The samples were cleaved at the same temperatures at which the ARPES measurements were performed (8 K and 120 K).
All the data have been recorded within 8 hours after the cleavage.
Within this time period, we did not observe discernible shifts in band energy.
Thus, the doping effect discussed later is not related to the time-dependent doping effect reported earlier \cite{LuoarXiv2021}.

\begin{figure}
\includegraphics[width=3.4in]{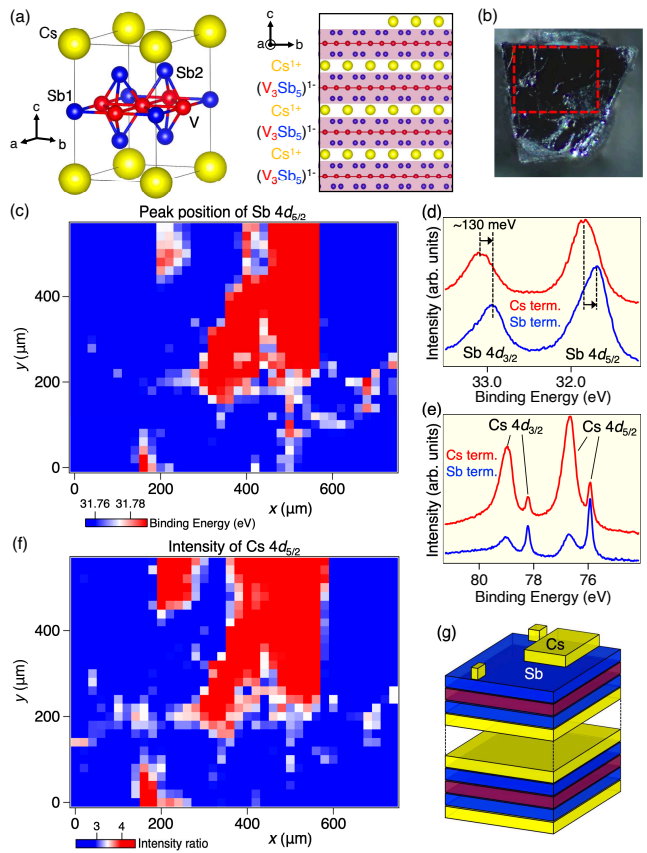}
\vspace{0cm}
\caption{
 (a) Crystal structure of $\rm{CsV}_{3}\rm{Sb}_{5}$.
 (b) Photograph of a cleaved $\rm{CsV}_{3}\rm{Sb}_{5}$ single crystal.
 (c) Spatial map of the $E_{\rm{B}}$ position of Sb-$4d_{5/2}$ core level measured with $h\nu$ = 106 eV at $T$ = 8 K in the spatial region enclosed by red dashed rectangle (area size: 600$\times$800 $\mu$m$^{2}$) in (b), obtained with the step size of 20 $\mu$m.
 (d), (e) EDC in the Sb-$4d$ and Cs-$4d$ core-level regions, respectively, for the Cs-(red) and Sb-(blue) terminated surfaces.
 (f) Spatial map of the intensity ratio of the surface/bulk Cs-$4d_{5/2}$ core-level peaks.
 (g) Schematic view of the cleaved surface of $\rm{CsV}_{3}\rm{Sb}_{5}$.
}
\end{figure}

First, we demonstrate how to distinguish two types of polar surfaces.
The cleaved surface of $\rm{CsV}_{3}\rm{Sb}_{5}$ contains a flat shiny mirror-like region [Fig. 1(b)].
By sweeping the micro-focused photon beam on the surface, we have mapped out the spatial distribution of the binding energy ($E_{\rm{B}}$) of Sb-$4d_{5/2}$ core-level peak [see Fig. 1(c)], which is a good measure of the surface doping level (note that the selection of Sb-$4d_{3/2}$ peak reproduces the same result).
As seen in Fig. 1(c), the surface looks to consist of two regions (blue and red regions).
The core-level spectrum in the red region [red curve in Fig. 1(d)] is shifted as a whole toward higher $E_{\rm{B}}$ by $\sim$130 meV with respect to that in the blue region (blue curve), indicating that the red region is more electron doped than the blue region (see Section 1 of Supplemental Material for detailed discussion on the origin of the Sb-$4d$ core-level shift \cite{refSM}).
Such a difference in the doping level signifies different polarities of the two regions.
Here we assign the electron (hole)-rich red (blue) region to the Cs (Sb)-terminated surface because Cs atoms on top surface do not need to donate electrons to the missing upper V$_3$Sb$_5$ layer which is removed by the cleaving, so that the V$_3$Sb$_5$ layer beneath the topmost Cs layer is more electron doped for the Cs-terminated surface. Inversely, the V$_3$Sb$_5$ layer at the surface is hole-doped compared with the bulk counterpart.
This is also corroborated by the behavior of the Cs-$4d$ core level.
Figure 1(e) shows the Cs-$4d$ core-level spectra measured at the same spatial location as in Fig. 1(d), where we observe that each spin-orbit satellite of Cs-$4d$ core levels ($d_{5/2}$ or $d_{3/2}$) consists of two peaks; a peak at lower $E_{\rm{B}}$ originates from bulk Cs atoms whereas that at higher $E_{\rm{B}}$ is from surface Cs atoms, as confirmed by the angular dependence of their intensity ratio (Section 1 of Supplemental Material \cite{refSM}).
One can recognize that the surface Cs peak is stronger than the bulk one for the Cs-termination-dominant surface (red curve), whereas it is weaker for the Sb-termination-dominant case (blue curve) (see Section 2 of Supplemental Material for the analysis of peak position \cite{refSM}).
The spatial image of the intensity ratio between the surface and bulk Cs peaks in Fig. 1(f) shows a good agreement with that of the Sb-core-level $E_{\rm{B}}$ in Fig. 1(c).
This supports our assignment of the Cs- and Sb-termination-dominant surfaces (hereafter, we simply call them Cs- and Sb-terminated surfaces, respectively), as schematically shown in Fig. 1(g).

We show in Figs. 2(a) and 2(b) the ARPES intensity at $T$ = 120 K (above $T_{\rm{CDW}}$ = 93 K) along the $\Gamma$KM cut ($k_{z}\sim0$) [red line in Fig. 2(c)] for the Cs- and Sb-terminated surfaces, respectively.
One can recognize overall similarity in the band structure between the two, \textit{i.e.} an electron band at the $\Gamma$ point ($\alpha$) with the Sb-$5p_{z}$ character, linearly dispersive V-$3d_{xz/yz}$ bands ($\beta,\gamma$) forming a Dirac cone near $E_{\rm{F}}$ (marked by an arrow), the $\delta$ band with the V-$3d_{xy/x^{2}-y^{2}}$ character forming a saddle point (SP) slightly above $E_{\rm{F}}$, and the $\varepsilon$ band that intersects the $\delta$ band to form a Dirac point at the K point at $E_{\rm{B}}\sim0.3$ eV \cite{NakayamaPRB2021,NakayamaPRX2022,JiangNM2021,WangPRB2022,TanPRL2021,FuPRL2021,ZhaoNature2021}.
A closer look further reveals some intrinsic differences; the band structure of the Cs-terminated surface is shifted downward as a whole with respect to that of Sb-terminated surface, as recognized by a direct comparison of surface-termination-dependent ARPES spectra at representative $\textbf{k}$ points plotted in Figs. 2(d)-2(f), in which numerical fittings to the peak position clarified the energy shift to be 30-90 meV (note that a finite variation in the magnitude of the energy shift among different bands and Sb-$4d$ core levels suggests a non-rigid-band-type energy shift; also see Section 3 of Supplemental Material for the energy shift of the $\delta$ band \cite{refSM}).
Moreover, while the Sb-terminated surface exhibits a single $\alpha$ band bottomed at $E_{\rm{B}}\sim0.5$ eV, it splits into two bands bottomed at $E_{\rm{B}}\sim0.5$ and 0.8 eV for the Cs-terminated surface.
This suggests that not only polarity but also the band structure above $T_{\rm{CDW}}$ is different between the Cs- and Sb-terminated surfaces.
We have performed ARPES measurements more than three times with different samples for each surface termination and confirmed the reproducibility (see Section 4 of Supplemental Material \cite{refSM}).

\begin{figure}
\includegraphics[width=3.4in]{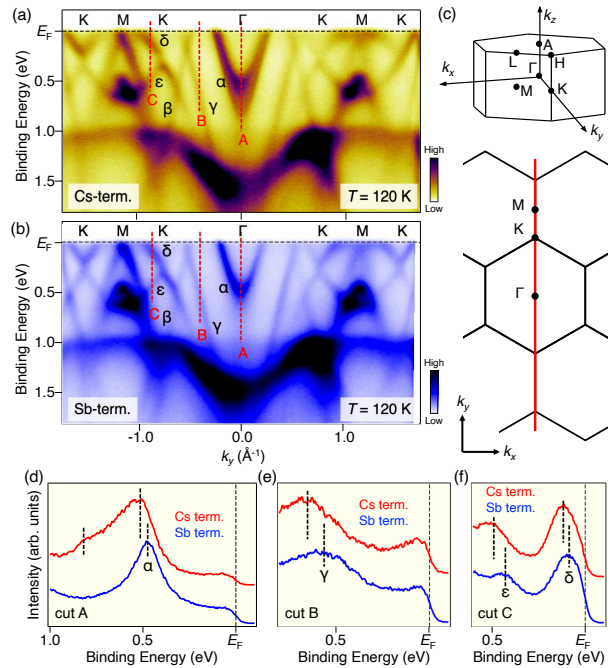}
\vspace{0cm}
\caption{
(a), (b) ARPES intensity at $T$ = 120 K for Cs- and Sb-terminated surfaces, respectively, measured with $h\nu$ = 106 eV along the $\Gamma$KM cut.
(c) (top) Bulk Brillouin zone and (bottom) slice of Brillouin zone at $k_z$ = 0 plane.
(d)-(f) Comparison of the EDCs measured for the Cs (red) and Sb (blue) surface terminations at \textbf{k} cuts A-C, respectively.
The position of \textbf{k} cuts is indicated by red dashed lines in (a) and (b).
}
\end{figure}

\begin{figure*}
\includegraphics[width=6.8in]{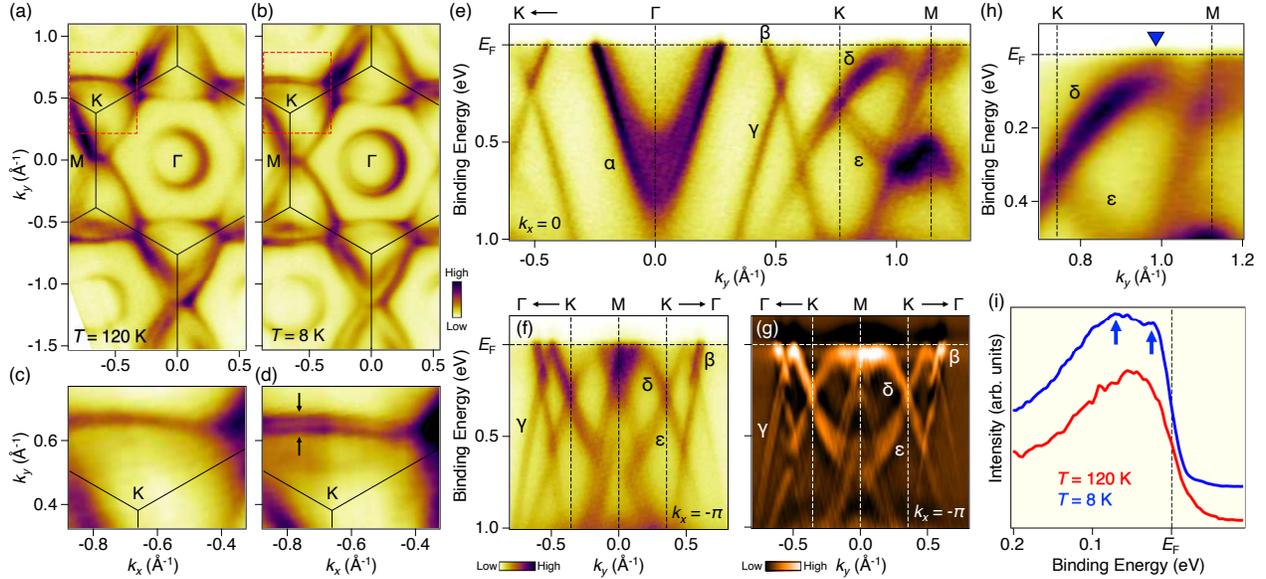}
\vspace{0cm}
\caption{
(a), (b) FS map for the Cs-terminated surface measured at $T$ = 120 K and 8 K, respectively.
(c), (d) Enlarged view in the $\textbf{k}$ region enclosed by dashed red rectangle in (a) and (b), respectively.
(e), (f) ARPES intensity plots at $T$ = 8 K measured along $k_{y}$ at $k_{x}=0$ ($\Gamma$KM cut) and $-\pi$ (KMK cut), respectively.
(g) Second-derivative intensity of (f).
(h) Plot of near-$E_{\rm{F}}$ ARPES intensity at $T$ = 8 K along the MK cut obtained with higher resolution.
(i) EDCs at $T$ = 120 K (red) and 8 K (blue) at the $k_{\rm{F}}$ point along the MK cut [blue triangle in (h)].
The hump and peak structures at $T$ = 8 K are indicated by blue arrows.
}
\end{figure*}

Next we analyze the electronic structure of the Cs-terminated surface in more detail, particularly in relation to the CDW transition.
Figures 3(a) and 3(b) show a comparison of the Fermi-surface (FS) mapping at the $k_{z}\sim0$ plane between $T$ = 120 K (above $T_{\rm{CDW}}$) and 8 K (below $T_{\rm{CDW}}$).
At $T$ = 120 K, one can recognize a small circular and a large hexagonal pocket centered at the $\Gamma$ point, together with a triangular pocket at the K point \cite{LiPRX2021,LouPRL2022,NakayamaPRB2021,LiuPRX2021,ChoPRL2021,LuoNCOM2022,KangNP2022,KatoCOMMAT2022,HuNCOM2022}.
On the other hand, at $T$ = 8 K, the hexagonal pocket splits into two pockets as more clearly seen in Figs. 3(c) and 3(d), where the enlarged FS image is displayed.
Such a band splitting is observed in a wide ($E,\textbf{k}$) region, as highlighted by the ARPES intensity at $T$ = 8 K measured along the representative $k_{y}$ cuts at $k_{x}$ = 0 and $-\pi$ in Figs. 3(e) and 3(f).
For example, the $\delta$ and the $\varepsilon$ bands show an energy splitting as better visualized in Fig. 3(e) [also see the energy distribution curve (EDC) in Sec. 5 of the Supplemental Material [35]]. Also, the $\beta$ and $\gamma$ bands show splitting as seen in Fig. 3(f) [note that there are two peaks in the momentum distribution curve [(MDC); red curves] in the Supplemental Material [35]] and in the corresponding second-derivative intensity in Fig. 3(g).
Thus, in the CDW phase, all the V-derived bands show splitting.
It is noted, however, that although the Sb-derived $\alpha$ band appears to split, it is unclear whether it is of the CDW origin, because the splitting was also observed at $T$ = 120 K [Fig. 2(a)].
The origin of the $\alpha$-band splitting in the normal state has been discussed in connection with the formation of quantum well states \cite{CaiarXiv2021} and the $k_z$ broadening effect \cite{NakayamaPRB2021}.

Now, we discuss the origin of the band splitting for the V-derived bands.
Apparently, it is not associated with the exchange splitting, because no signature of ferromagnetism was reported in $\rm{AV}_{3}\rm{Sb}_{5}$ \cite{OrtizPRM2019}.
The existence of surface states near the main bulk bands and the influence of the $k_{z}$ broadening would be also unlikely, because the splitting is absent at $T$ = 120 K.
A plausible explanation is the CDW-potential-induced band folding along the $k_{z}$ direction, which is naturally expected from the reported 3D nature of CDW accompanied by the out-of-plane unit-cell doubling (2$\times$2$\times$2) \cite{LiangPRX2021,LiPRX2021} (see Fig. S8 in Supplemental Material for details \cite{refSM}).
It is noted that we also found a signature of the in-plane band folding, as detailed in Section 6 of Supplemental Material \cite{refSM}.

We measured the $\delta$ band dispersion along the MK cut which forms the SP at $T$ = 8 K [Fig. 3(h)] and revealed that it disperses back toward higher $E_{\rm{B}}$ without reaching $E_{\rm{F}}$ due to the CDW-gap opening.
This gap is well seen from the EDC at the Fermi wave vector ($k_{\rm{F}}$) [indicated by triangle in Fig. 3(h)] at $T$ = 8 K, where a characteristic hump structure appears at $E_{\rm{B}}\sim70$ meV in agreement with previous studies \cite{NakayamaPRB2021,LiuPRX2021}.
Besides the 70-meV hump, the EDC at $T$ = 8 K shows a small peak at $E_{\rm{B}}\sim20$ meV, which may be due to the CDW-gap opening in the folded band from the $k_{z}=\pi$ plane.

\begin{figure}
\includegraphics[width=3.4in]{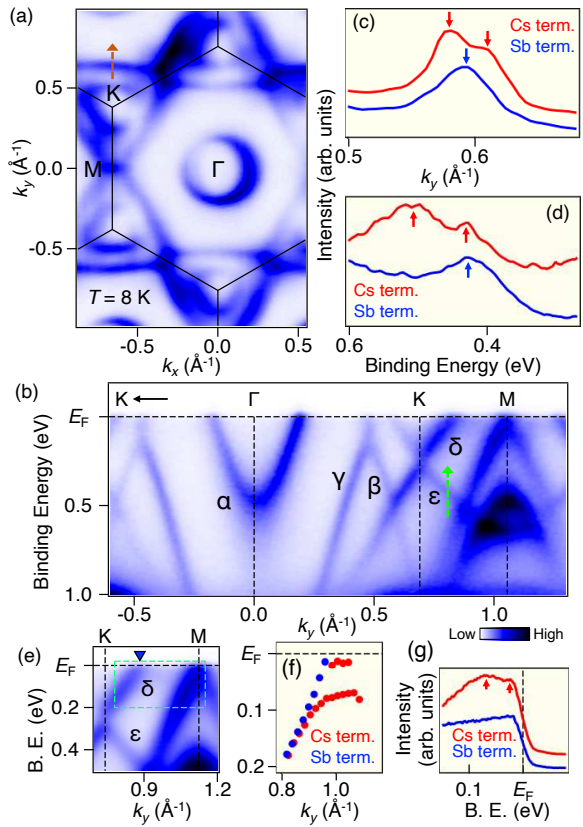}
\vspace{0cm}
\caption{
(a) FS map for the Sb-terminated surface measured at $T$ = 8 K with $h\nu$ = 106 eV.
(b) ARPES-intensity plot at $T$ = 8 K measured along the $\Gamma$KM cut.
(c) Comparison of the MDCs between the Sb- and Cs-terminated surfaces, obtained along the $\textbf{k}$ cut indicated by brown dashed arrow in (a).
(d) Comparison of EDC between the two surface terminations, obtained in the $E_{\rm{B}}$ region shown by dashed green arrow in (b).
Arrows in (c) and (d) indicate peak positions of the MDC/EDC.
(e) Plot of near-$E_{\rm{F}}$ ARPES intensity for the Sb-terminated surface, measured along the MK cut with higher energy resolution at $T$ = 8 K.
(f) Comparison of experimental band dispersions between the Sb- and Cs-terminated surfaces in the ($E, \textbf{k}$) region enclosed by dashed green rectangle in (e).
(g) Comparison of EDC at the $k_{\rm{F}}$ point of the $\delta$ band [blue triangle in (e)].
Red arrows show hump-and-peak structure associated with the CDW gap for the Cs-terminated surface.
}
\end{figure}

Having established the electronic states of the Cs-terminated surface, now we turn our attention to the Sb-terminated surface.
Figure 4(a) shows the typical FS mapping for the Sb-terminated surface at $T$ = 8 K well below $T_{\rm{CDW}}$.
By comparing with that of the Cs-terminated counterpart [Fig. 3(b)], we observe shrinkage of the circular pocket centered at the $\Gamma$ point and the triangular pocket at the K point due to the excess hole doping in the Sb-terminated surface (see Section 7 of Supplemental Material \cite{refSM} for quantitative comparison).
All the energy bands observed in this study have a surface character (specifically, the energy bands of the topmost V$_3$Sb$_5$ layer) because of a relatively short escape depth of photoelectrons ($\sim$10 \AA).
For the Sb-terminated surface, the hole-doped character of the $\Gamma$-centered pocket derived from the Sb1 atoms suggests that the hole-doping effect extends to the Sb1-V plane.
Taking into account these points, it is natural to observe an energy shift of the V-derived bands.
It is noted that, besides the polarity-induced charge doping, the inversion symmetry is broken at the Sb-terminated surface.
But the band splitting, e.g. Rashba-type spin splitting, expected by the inversion-symmetry breaking has not been observed within our experimental uncertainty, suggesting that such an effect may not be strong in CsV$_3$Sb$_5$.

More importantly, the band dispersion is basically temperature independent as seen from the ARPES intensity at $T$ = 8 K in Fig. 4(b) which is similar to that at $T$ = 120 K [Fig. 2(b)]; one can see no doubling of the Sb and V-derived bands.
The absence of band doubling for the Sb-terminated surface is more clearly visualized in Figs. 4(c) and 4(d), where the representative MDCs and EDCs are compared between the Cs- and Sb-terminated surfaces (red and blue curves, respectively).
We also found that the Sb-terminated surface shows no clear CDW-gap opening.
As shown in Fig. 4(e), the $\delta$ band simply crosses $E_{\rm{F}}$ (see triangle) without bending back behavior, in contrast to the case of the Cs-terminated surface [see the band dispersion extracted from the peak position of EDCs in Fig. 4(f)].
Also, the EDC at the $k_{\rm{F}}$ point for the Sb-terminated surface [blue curve in Fig. 4(g)] exhibits no hump structure unlike the case of the Cs-terminated counterpart (red curve).

The band diagram as a function of surface Cs coverage derived from the present study and our previous work on Cs-dosed CsV$_3$Sb$_5$ \cite{NakayamaPRX2022} is summarized in Fig. 5. The excess hole doping realized in the Sb-terminated surface pushes up the energy bands, whereas the electron doping in the Cs-terminated or Cs-dosed surfaces pushes down (note that $E_{\rm F}$ for the bulk band structure would be obtained by averaging $E_{\rm F}$ of the Cs and Sb terminations because the electron concentration identical to the bulk is realized at the Cs coverage of 50 $\%$). Despite the surface character of observed band structures in the present study, a signature of 2$\times$2$\times$2 CDW is found in the Cs-terminated surface at low temperatures, suggesting that the 2$\times$2$\times$2 CDW and the resultant band folding along $k_z$ can survive even at the surface under a certain condition. On the other hand, the 2$\times$2$\times$2 CDW is suppressed in the Sb-terminated and heavily Cs-dosed surfaces. Understanding the relationship between this contrasting CDW behavior and the difference in surface charge is a next important step toward clarifying the mechanism of unconventional CDW in CsV$_3$Sb$_5$.

\begin{figure}
\includegraphics[width=3.4in]{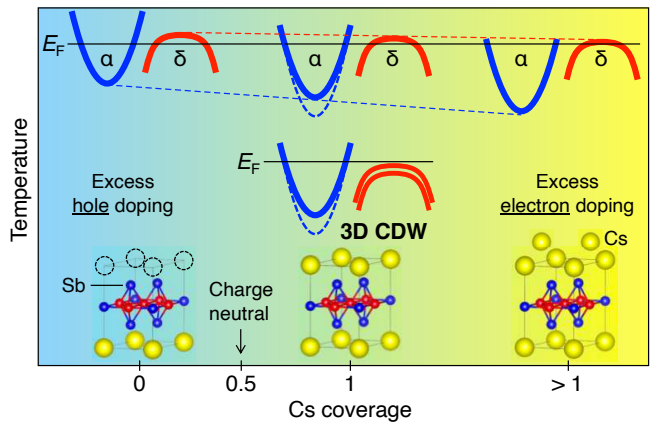}
\vspace{0cm}
\caption{
Schematic of key band structure in CsV$_3$Sb$_5$ as a function of surface Cs coverage.
Inset shows the crystal structure at the surface of Sb-terminated (left), Cs-terminated (middle), and Cs-dosed $\rm{CsV}_{3}\rm{Sb}_{5}$ (right).
}
\end{figure}

We discuss implications of the present results in relation to the exotic surface properties recently reported by STM of $\rm{AV}_{3}\rm{Sb}_{5}$ \cite{ChenNature2021,ZhaoNature2021,LiangPRX2021,WangPRB2021,ShumiyaPRB2021,NieNature2022,LiarXiv2021}.
One of the intriguing surface properties is that the superconducting-gap structure shows a higher density of states at $E_{\rm{F}}$ for the Cs-terminated surface compared to the Sb-terminated one \cite{ChenNature2021}.
This could be interpreted in terms of the enhanced gap anisotropy in the Cs-terminated surface.
In light of our observations, such a difference is suggested to be related to the strength of CDW at the surface; namely, the Cs-terminated surface is more strongly perturbed by CDW so that the superconducting-gap structure becomes more complex and unconventional.
Another exotic surface property is the emergence of surface charge order with a unidirectional 4$\times$1 periodicity that appears only for the Sb-terminated surface \cite{ChenNature2021,ZhaoNature2021,LiangPRX2021,WangPRB2021,ShumiyaPRB2021,NieNature2022,LiarXiv2021}.
This surface charge order may be associated with the suppression of the 2$\times$2 CDW in light of the present ARPES result (in fact, one may see a finite change in the intensity of 2$\times$2 superspots in the Fourier transform image of the STM pattern \cite{LiangPRX2021}).
Also, the mechanism of the 4$\times$1 charge order cannot be easily explained from the observed FS topology; it is hard to infer from the FS mapping in Figs. 3(b) and 4(a) that the Sb-terminated surface possesses a good nesting condition with the 4$\times$1 periodicity.
This suggests that other possibilities beyond the simple FS nesting such as electron correlation and surface instability \cite{WangPRB2021,ChenNature2021,ZhaoNature2021} must be taken into account to understand the unconventional surface charge order; we leave this issue as an open question.
In any case, the present study strongly suggests that the polarity-dependent band characteristics, such as the critical difference in the carrier doping levels and CDW properties, must be properly taken into account to understand the microscopic mechanism behind the intriguing surface electronic anomalies in $\rm{AV}_{3}\rm{Sb}_{5}$.

In conclusion, the present micro-ARPES study of $\rm{CsV}_{3}\rm{Sb}_{5}$ has revealed the polar nature of the cleaved surface and self-doping effects.
We uncovered that the Sb-terminated surface is characterized by excess hole doping and resultant suppression of CDW, whereas the Cs-terminated surface shows the band doubling and the resultant CDW-gap opening below $T_{\rm{CDW}}$ which support the 3D CDW.
The present result opens a pathway toward understanding and manipulating exotic properties in kagome superconductors $\rm{AV}_{3}\rm{Sb}_{5}$.

\begin{acknowledgments}
This work was supported by JST-CREST (No. JPMJCR18T1), JST-PRESTO (No. JPMJPR18L7), Grant-in-Aid for Scientific Research (JSPS KAKENHI Grant Numbers JP21H04435 and JP20H01847), KEK-PF (Proposal number: 2021S2-001), and the Sasakawa Scientific Research Grant from the Japan Science Society.
The work at Beijing was supported by the National Key R\&D Program of China Grant No. 2020YFA0308800), the Natural Science Foundation of China (Grants No. 92065109), the Beijing Natural Science Foundation (Grant No. Z210006), and the Beijing Institute of Technology (BIT) Research Fund Program for Young Scholars (Grant No. 3180012222011).
T.K. acknowledges support from GP-Spin at Tohoku University and JST-SPRING (No. JPMJSP2114).
Z.W. thanks the Analysis \& Testing Center at BIT for assistance in facility support.
\end{acknowledgments}

\bibliographystyle{prsty}

\section{SUPPLEMENTAL MATERIAL}
\subsection{Section 1: Photoelectron emission angle dependence of core-level spectra}
Figure 6 shows a comparison of core-level spectra measured at different photoelectron emission angles $\theta = 90^\circ$ and $60^\circ$ [see Fig. 6(a) for the definition of $\theta$].
Because of the finite escape depth of photoelectrons ($\lambda$) in the sample, the photoemission spectrum taken at a smaller $\theta$ is more surface sensitive; an effective escape depth $\lambda_{\rm eff}=\lambda\sin{\theta}$ is reduced by $\sim$15\% by changing $\theta$ from $90^\circ$ to $60^{\circ}$ [see Fig. 6(a)].
If the Sb-$4d$ core levels had surface and bulk components at higher and lower binding energies ($E_{\rm{B}}$’s), respectively, and the surface component became stronger in intensity by changing $\theta$ from $90^\circ$ to $60^\circ$ as observed for the Cs-$4d$ core levels (see below), the peak position of Sb-$4d$ core levels at $\theta = 60^\circ$ would be located at higher $E_{\rm{B}}$ than that at $\theta = 90^\circ$.
However, as shown in Fig. 6(b), such an energy shift is not seen within the present experimental uncertainty of 10 meV, which is much smaller than the 130-meV energy shift observed by the termination-dependent ARPES study.
This result is not consistent with the two-components scenario, but supports the doping scenario as the origin of the energy shift in the Sb-$4d$ core level [Fig. 1(d)].
In contrast, the Cs-$4d$ core levels show surface and bulk components.
As shown in Fig. 6(c), each spin-orbit satellite of Cs-$4d$ core levels ($d_{\rm{5/2}}$ or $d_{\rm{3/2}}$) consists of two peaks.
A peak at lower $E_{\rm B}$ originates from bulk Cs atoms, whereas that at higher $E_{\rm B}$ is from surface Cs atoms, as supported by the angular dependence of their intensity ratio; the shorter $\lambda_{\rm eff}$ at $\theta = 60^{\circ}$ leads to the stronger surface-derived (weaker bulk-derived) peak intensity of the Cs-$4d$ core levels.

\begin{figure}
\includegraphics[width=3.4in]{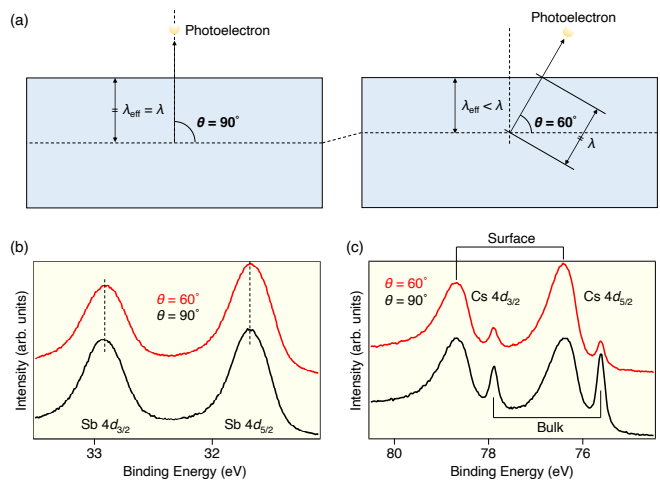}
 \caption{
(a) Schematic of experimental setup and the definition of photoelectron emission angle $\theta$, escape depth $\lambda$, and effective escape depth $\lambda_{\rm eff}$. (b) Comparison of EDCs in the Sb-$4d$ core-level region measured at $\theta = 90^{\circ}$ (black) and $60^{\circ}$ (red) for the Cs-terminated surface. (c) Same as (b) but for the Cs-$4d$ core-level region.
}
\end{figure}

\begin{figure}
\includegraphics[width=2in]{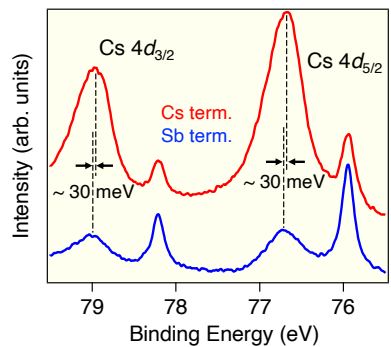}
 \caption{
Comparison of energy distribution curves (EDCs) in the Cs-$4d$ core-level region measured for Cs- and Sb-terminated surfaces (red and blue curves, respectively).
}
\end{figure}

\subsection{Section 2: Peak position of the Cs-$4d$ core levels}
The termination-dependent Cs-$4d$ core-level data in Fig. 7 shows that the surface components of the Cs-terminated surface shift toward lower $E_{\rm{B}}$ by $\sim$30 meV with respect to those of the Sb-terminated surface.
Since this direction is opposite from that observed in the Sb-$4d$ core levels, the observed Cs-$4d$ core-level shift cannot be understood in terms of the carrier doping, but probably reflects a difference in the density of Cs atoms at the surface.
It is known that the core level of alkaline-metal atoms shifts toward lower $E_{\rm{B}}$ with increasing the number of alkaline-metal atoms per area.
Such a characteristic energy shift is theoretically explained by a Born-Haber cycle \cite{PirugSurfSci1985} and provides a reasonable explanation on our observation.
Namely, the Cs-terminated surface has the higher density of Cs atoms compared to that of the Sb-terminated surface and hence the Cs-$4d$ core level appears at lower $E_{\rm{B}}$, whereas the Sb-$4d$ core levels shift toward higher $E_{\rm{B}}$ because of the doping-induced chemical potential shift.
A similar behavior was reported in K-dosed Fe(Se,Te) \cite{ZhangAPL2014} in which gradual K deposition on the surface leads to the shift of the Te-$4d$ core level toward higher $E_{\rm{B}}$ due to the electron doping to the Fe(Se,Te) layer, whereas the K-$3p$ core level shifts toward lower $E_{\rm{B}}$.

\begin{figure}
\includegraphics[width=3.4in]{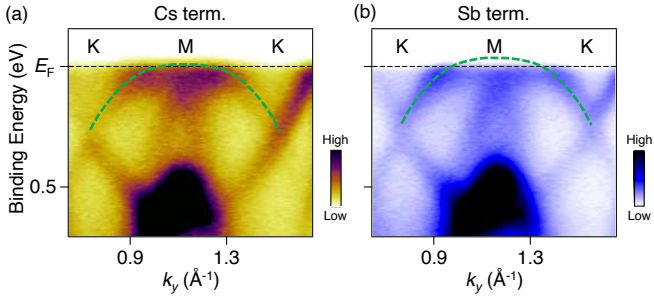}
 \caption{
(a), (b) ARPES intensity measured along the KM cut at $T = 120$ K in the Cs- and Sb-terminated surfaces, respectively.
Green dashed curves are a guide for the eyes to trace the SP-band dispersions.
}
\end{figure}

\subsection{Section 3: Surface-termination dependence of the saddle-point position}
Figure 8 shows the ARPES data that supports the SP-band energy shift, in which a comparison of the ARPES intensity along the KMK cut between the Cs- and Sb-terminated surfaces is made.
One can directly recognize that the saddle point (SP) band (green dashed curve) is located at lower $E_{\rm{B}}$ in the Sb-terminated surface compared to that in the Cs-terminated surface, signifying the difference in the doping level.

\subsection{Section 4: Reproducibility of polar-surface formation}
We have performed ARPES measurements more than three times with different samples for each surface termination and confirmed the reproducibility, e.g. the termination-dependent band energy shifts, as seen by a comparison of the band dispersions measured with different samples in Fig. 9.
We also performed ARPES measurements with some different photon energies between 70 and 130 eV which have comparable probing depths, and obtained qualitatively the same results.

\subsection{Section 5: Band splitting observed in the Cs-terminated surface}
Figure 10 shows representative EDCs and momentum distribution curves (MDCs) at $T = 8$ K measured for the Cs-terminated surface.
One can clearly see a two-peaked structure in EDCs [Figs. 10(c) and 10(d)] and MDCs [Figs. 10(e) and 10(f)] which supports the band splitting associated with the CDW-induced band folding.

\begin{figure}
\includegraphics[width=3.4in]{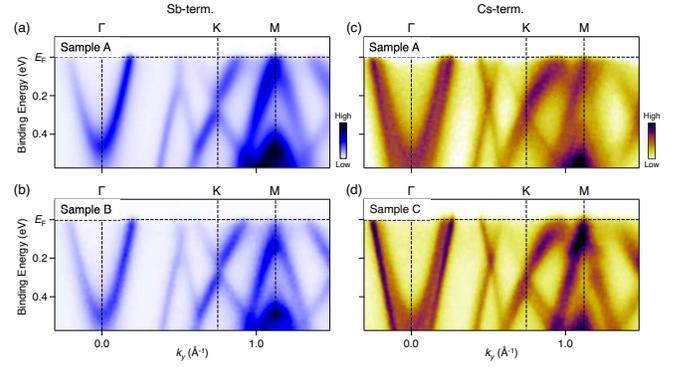}
 \caption{
(a), (b) Comparison of the band dispersions at $T = 8$ K for the Sb-terminated surface between two different $\rm{CsV}_{3}\rm{Sb}_{5}$ samples (samples A and B). (c), (d) Same as (a) and (b) but for the Cs-terminated surface.}
\end{figure}

\subsection{Section 6: A signature of in-plane band folding}
Figures 11(a) and 11(b) show a comparison of the second-derivative ARPES intensity along the $\Gamma$KMK$\Gamma$ cut at $T = 8$ K and 120 K for the Cs-terminated surface.
We observed the discontinuity of the $\beta$ band dispersion along the $\Gamma$K line of the Cs-terminated surface at $T = 8$ K, as indicated by white arrows in Fig. 11(a).
This anomaly is of CDW origin because it disappears at $T = 120$ K [Fig. 11(b)] and is well reproduced by the density-functional theory that incorporates only the in-plane $2\times2$ periodicity [Fig. 11(c)].
Therefore, the Cs-terminated surface shows the unit-cell doubling not only along the $c$ axis but also in the $a-b$ plane.
On the other hand, for the Sb-terminated surface, the discontinuity of the $\beta$ band is absent even at 8 K, so both the c-axis and in-plane unit-cell doublings are suppressed.

\begin{figure}
\includegraphics[width=3.4in]{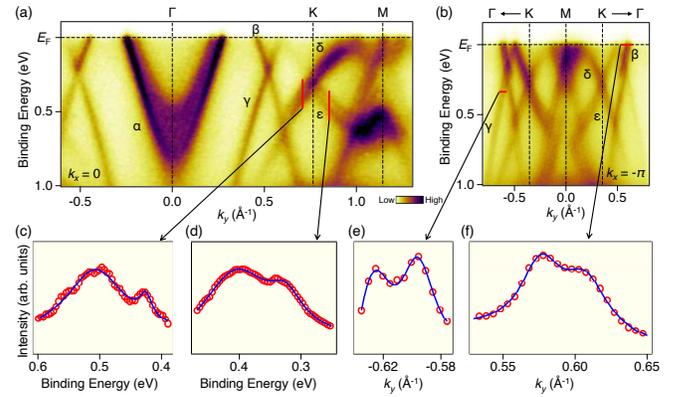}
 \caption{
(a), (b) ARPES intensity plots at $T = 8$ K measured along $k_{y}$ at $k_{y}=0$ ($\Gamma$KM cut) and $-\pi$ (KMK cut), respectively.
(c), (d) EDCs at representative energy slices indicated by red lines in (a).
(e), (f) MDCs at representative $\textbf{k}$ slices shown by red lines in (b).
Blue curves in (c)-(f) are the result of numerical fittings.
}
\end{figure}

\subsection{Section 7: Comparison of Fermi surface between Cs- and Sb-terminated surfaces}
Figures 12(a) and 12(b) show a comparison of the ARPES intensity map at the Fermi level ($E_{\rm{F}}$) around the K point between the Cs- and Sb-terminated surfaces. 
The peak position of MDC at $E_{\rm{F}}$ is overlaid for the triangular-shaped electron Fermi surface (FS; black circles and red triangles, respectively).
Direct comparison of the MDC peak position in Fig. 12(c) shows the shrinkage of the triangular FS in the Sb-terminated surface, consistent with the hole-doped character.

\begin{figure}
\includegraphics[width=3.4in]{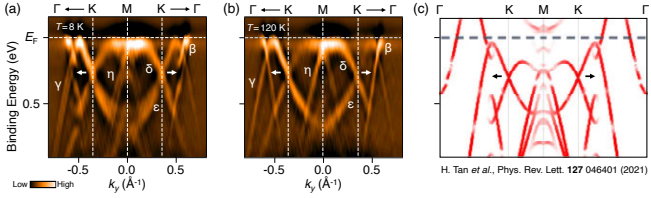}
 \caption{
(a), (b) Second-derivative ARPES-intensity plots for the Cs-terminated surface at $T = 8$ K and 120 K, respectively.
White arrows in (a) highlight the appearance of band hybridization by the in-plane band folding.
(c) Calculated band structure of $\rm{CsV}_{3}\rm{Sb}_{5}$ with the $2\times2\times1$ unit cell, reproduced from \cite{TanPRL2021}.
}
\end{figure}

\begin{figure}
\includegraphics[width=3.4in]{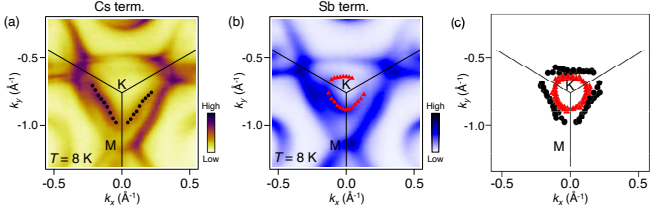}
 \caption{
(a), (b) ARPES intensity plots around the K point measured at $T = 8$ K for the Cs- and Sb-terminated surfaces, respectively.
Black and red triangles in (a) and (b) represent the Fermi wave vectors determined by numerical fittings to the MDCs at $E_{\rm{F}}$.
(c) Comparison of the Fermi wave vectors between the Cs- and Sb-terminated surfaces. The data points were symmetrized by assuming the $C_{3}$ symmetry centered at the K point.
}
\end{figure}

\begin{figure}
\includegraphics[width=3.4in]{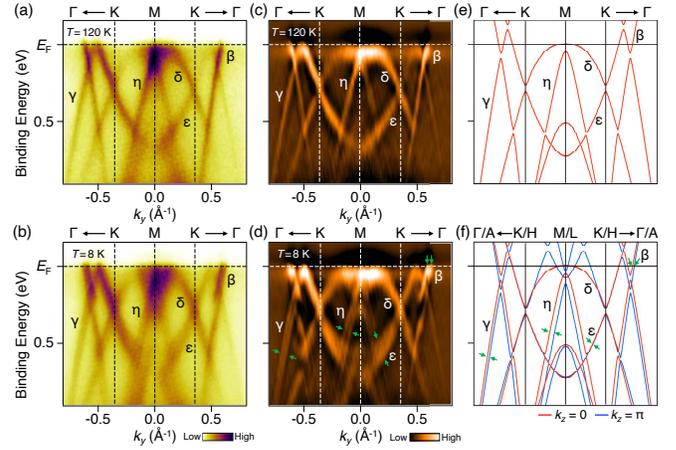}
 \caption{
(a), (b) ARPES-intensity plots at $T = 120$ K and 8 K, respectively, measured along the KM cut.
(c), (d) Second-derivative ARPES-intensity plots of (a) and (b), respectively.
(e), (f) Calculated band structures with SOC for (e) $k_{z} = 0$ (red) and (f) $k_{z} = 0$ (red) and $\pi$ (blue) (reproduced from \cite{NakayamaPRB2021}).
The calculations were carried out using the full-potential linearized augmented plane-wave method implemented in the WIEN2K code with generalized gradient approximation (GGA) and the Perdew-Burke-Ernzerhof (PBE) type exchange-correlation potential.
Spin-orbit coupling was included self-consistently, while the lattice parameters were directly obtained from experiments.
The $k$-points mesh for the irreducible Brillouin zone is $17\times17\times10$.
The Muffin-tin radii are 2.50 a.u. for Cs and V and 2.60 a.u. for Sb, respectively.
The maximum modulus for the reciprocal vectors K$_{\rm{max}}$ was chosen to satisfy $R_{\rm{MT}}\times\rm{K}_{\rm{max}}=8.0$.
}
\end{figure}

\subsection{Section 8: Comparison between the experimental and calculated band structures}
To clarify the origin of band doubling in the charge-density-wave (CDW) phase for the Cs-terminated surface of $\rm{CsV}_{3}\rm{Sb}_{5}$, we compare the ARPES results with the first-principles band-structure calculations. Figures 13(a) and 13(b) show the ARPES intensity along the KM cut at $T = 120$ K (above $T_{\rm{CDW}}$) and 8 K (below $T_{\rm{CDW}}$), respectively, measured with $h\nu = 106$ eV photons corresponding to the out-of-plane wave vector of $k_{z} \sim 0$ \cite{NakayamaPRB2021}.
The second-derivative intensity is displayed in Figs. 13(c) and 13(d).
At $T = 120$ K, we observe five dispersive bands (labelled as $\beta, \gamma, \delta, \varepsilon,$ and $\eta$) near $E_{\rm{F}}$ [Figs. 13(a) and 13(c)], in agreement with the first-principles calculations for the KM cut ($k_{z} = 0$) [Fig. 13(e)].
On the other hand, at $T = 8$ K, these bands show doubling as marked by green arrows in Fig. 13(d) [doubling of the $\delta$ band is seen in the data obtained with higher energy resolution in Figs. 3(h) and 3(i)].
This band doubling is explained in terms of the band folding along the $k_{z}$ direction due to the three-dimensional (3D) CDW \cite{LiangPRX2021}.
Namely, the doubling of unit-cell along the c-axis causes the folding of energy bands from $k_{z} = \pi$ to 0, and vice versa.
The calculated band dispersions superimposed with those of $k_{z} = 0$ and $\pi$ [Fig. 13(f)] is in good agreement with the experimental results, supporting the 3D-CDW origin of the band doubling for the Cs-terminated surface.
It is noted that to obtain a better matching between the experiment and calculation, additional level splitting associated with the repulsion between the original and folded bands (especially for the $\varepsilon$ band) as well as the CDW-gap opening near $E_{\rm{F}}$ needs to be taken into account.
It is also noted that, besides the $2\times2\times2$ periodicity, the $2\times2\times4$ periodicity has been also reported in $\rm{CsV}_{3}\rm{Sb}_{5}$ \cite{OrtizPRX2021}.
We naively expect the quadruple band splitting when the charge order has the $2\times2\times4$ periodicity.
But our data show only doubling. In this respect, our results are more consistent with the $2\times2\times2$ periodicity.

\end{document}